\documentclass[12pt,aps]{revtex4}
\usepackage{graphicx}

\clearpage

\begin{document}

\title{Strangeness Production Experiments at Jefferson Lab\footnote{Submitted to the 
Proceedings of ``SENDAI03: Electrophoto-production of Strangeness on Nucleons and 
Nuclei'', Tohoku University, Sendai, Japan, June 2003. To be published by World Scientific.
}}

\author{Reinhard A. Schumacher}

\affiliation{Department of Physics, Carnegie Mellon University, Pittsburgh, PA 15213, USA\\ 
E-mail: schumacher@cmu.edu}

\begin{abstract}
Experimental results for photo- and electro-production of open
strangeness from the Thomas Jefferson National Accelerator Facility
are discussed.  The results are from work completed by mid-2003 on
elementary $KY$ production, nuclear targets, and the exotic
$\Theta^+$ state.  It is shown how the increases in intensity and
precision of JLab experiments over earlier work have allowed new
phenomena to become measurable.

\end{abstract}

\maketitle

\section{Introduction}
Electro- and photo-production of open strangeness is interesting
because it involves a well-understood probe -- the photon --
interacting with a nucleon or nuclear target to make an $s\bar{s}$
quark pair that is not part of the target's valence structure.  This
makes reactions in which strange particles are produced attractive to
look for quark-level dynamics in the production mechanism.  Two
examples of possible quark-level dynamics are shown in this report.
Where direct signatures of quark degrees of freedom are not evident,
meson and baryon descriptions of the strangeness production mechanism
are sensitive to the inclusion of non-strange excited states of
nucleons that decay to strange particles in the final state.  For
example, excited non-strange baryons which are ``hidden'' when
searching in pionic final states may be ``revealed'' by looking in
final states with strangeness, simply because couplings to pion- and
kaon-containing final states will, in general, be different.  In the
area of strangeness production off nuclei, information about the
hyperon-nucleon interaction can be gleaned from final state
interactions with a deuteron target; a first result in this field has
been obtained.  The creation of hypernuclear states has been
demonstrated at Jefferson Lab for the first time, opening a new field
of research for the future.  The final topic in this report is the
recent evidence from Jefferson Lab of a strangeness +1 baryon dubbed
the $\Theta^+$.  At Jefferson Lab, this state has been seen with CLAS
in an exclusive reaction on the deuteron, as well as in a different
final state off the proton.

This report gives an overview of the experimental results that have
been obtained to date at Jefferson Lab in these fields. It also
mentions the experiments that are still under analysis or that have
yet to take data.  The sections we will use for the discussion are:

\begin{itemize}
\item {Elementary production} --- on the proton
\item {Nuclear electroproduction} --- on the deuteron and heavier nuclei
\item {The $\Theta^+$ state} --- first results from CLAS
\end{itemize}

\section{Elementary Production on the Proton}
The elementary photo- and electro-production of hyperons in the resonance
region is used mainly as a test of production mechanisms through
resonant baryon formation and kaon exchange.  Couplings of $N^*$'s and
$\Delta$'s to $KY$ final states differ from couplings to pions and
nucleons but are not well known.  Thus, one can expect to learn about
how the couplings compare to expectations (from flavor SU(3), for
instance), and possibly also to ``see'' resonance contributions that
are hidden in the case of pionic final states.

\begin{figure}[ht]
\vspace{-0.2in}
\resizebox{0.50\textwidth}{!}{\includegraphics{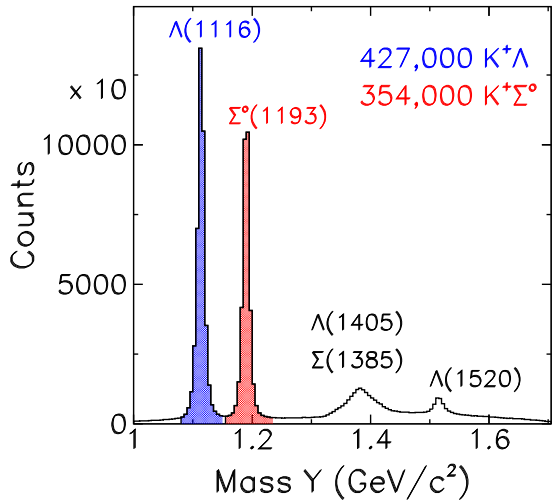}}   
\vspace{-0.2in}
\caption{Missing mass spectrum for photoproduction of strangeness in
CLAS, using a tagged bremsstrahlung beam with 2.4 GeV endpoint energy.
\label{fig:mmphoto}}
\end{figure}

\subsection{Photoproduction}\label{subsec:photo}
\begin{figure}[ht]
\resizebox{0.50\textwidth}{!}{\includegraphics{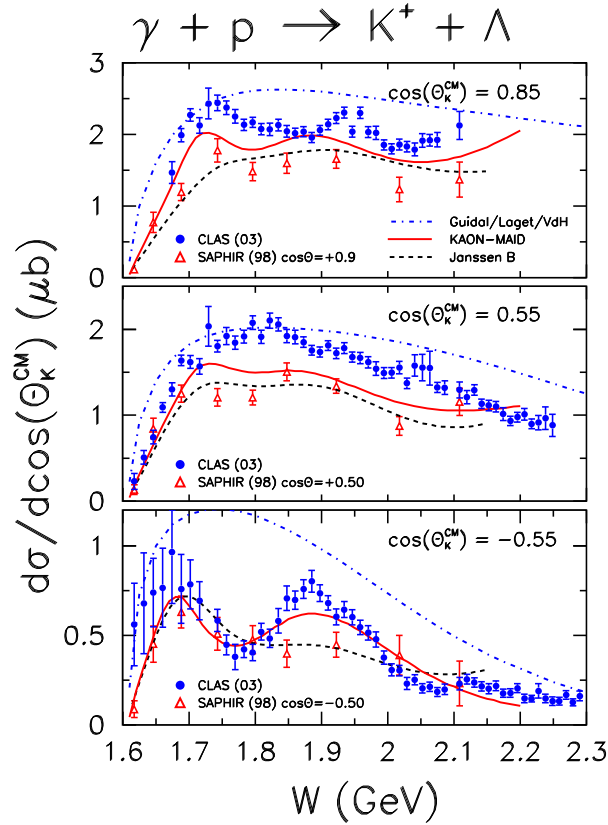}}   
\caption{
Energy dependence of the $\Lambda$ cross section at the most forward
angle measured (top), and at intermediate forward and backward angles
(middle, bottom); from McNabb {\it et al.}.~\cite{mcnabb}.  The curves
are for effective Lagrangian calculations computed by Kaon-MAID
~\cite{maid,mart} (solid), the related model of Janssen {\it et
al.}~\cite{jan} (dashed), and a Regge-model calculation of Guidal {\it
et al.} \cite{laget} (dot-dashed).  Data from SAPHIR~\cite{bonn} (open
triangles) are also shown.
\label{fig:b}}
\end{figure}

CLAS~\cite{clas0} has measured~\cite{mcnabb} the cross sections and
recoil polarizations for the reactions $\gamma + p \rightarrow K^+ +
\Lambda$ and $\gamma + p \rightarrow K^+ + \Sigma^0$ from threshold up
to 2.3 GeV.  The mass resolution for the detected kaons was $\sigma =
6.1$ MeV, as determined from time of flight.  This is shown in
Fig.~\ref{fig:mmphoto}.  One of the main results from this work is the
energy dependence of the $K^+\Lambda$ cross sections shown in
Fig.~\ref{fig:b}.  The panels show how, as a function of the
$s$-channel energy, $W$, the cross section evolves at very forward
(top), forward (middle), and backward (bottom) center-of-mass angles.
At the extreme angles there are peaks near threshold and near 1.9 GeV,
while in the middle-forward region the dip between the peaks is filled
in.  The higher-energy peaks are significant since they confirm hints
of this structure published by the SAPHIR
collaboration~\cite{bonn}. They indicate that there are at least two
structures in the mass range near 1.9 GeV which couple strongly to
$K^+\Lambda$, since the shape of the peaks and their position change
somewhat with angle.  Three calculations are shown for reference.  To
mention just one feature, the MAID model~\cite{maid} is based on the
hadrodynamic model of Mart {\it et al.}~\cite{mart}, and generates the
higher energy peak with a $D_{13}$(1895) state.  Clearly this is not
enough to describe the data.  These data will now allow more detailed
modeling of the production of $K^+\Lambda$ to be constructed.

The hyperon recoil polarization has also been measured, as can be seen
in the unpublished portion of Ref~\cite{mcnabb}.  CLAS also has data
under analysis of photoproduction using linearly and circularly
polarized photon beams.  The observables $\Sigma$ (linear polarized
beam asymmetry)~\cite{klein}, $C_x^\prime$ and $C_z^\prime$
(beam-recoil double polarizations)~\cite{bradford} will eventually be
available, measured over the same kinematic range as the results
mentioned above.

\subsection{Electroproduction}\label{subsec:electro}
\begin{figure}[ht]
\resizebox{0.50\textwidth}{!}{\includegraphics{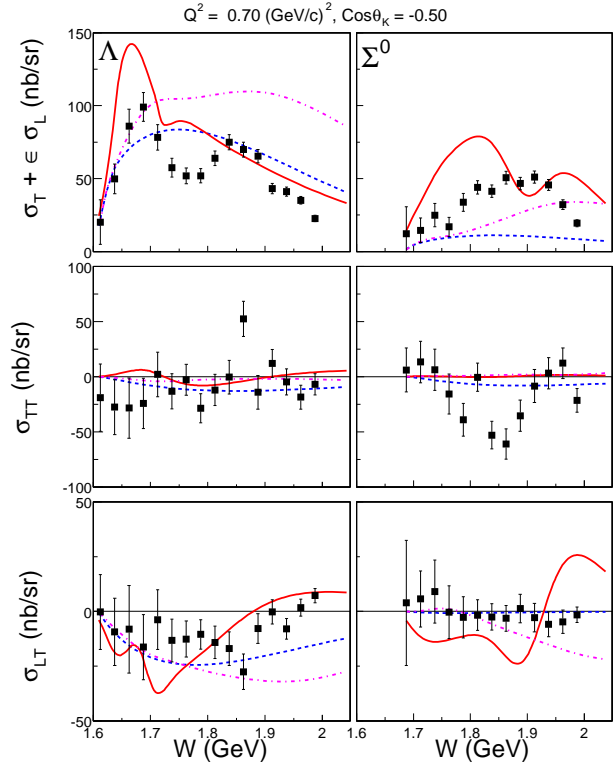}}   
\caption{
Energy dependence of the $\Lambda$ (left) and $\Sigma^0$ (right) cross
sections for $\sigma_U = \sigma_T + \epsilon\sigma_L$ (top),
$\sigma_{TT}$ (middle), and $\sigma_{LT}$ (bottom).  The curves are for
effective Lagrangian calculations computed by Kaon-MAID~\cite{maid}
(solid), the similar model of Janssen {\it et al.}~\cite{jan}
(dot-dashed), and a Regge-model calculation of Guidal {\it et
al.}~\cite{laget} (dashed).
\label{fig:c}}
\end{figure}

Moving away from the kinematic point at $Q^2=0$, electroproduction of
strangeness introduces two new ingredients into the discussion: the
longitudinal coupling of the photons in the initial state, and the
electromagnetic and hadronic form factors of the exchanged particles.
New observables include the longitudinal-transverse and
transverse-transverse interference terms in the cross section. These
have been measured by CLAS for the first time over a broad kinematic
range.  The expression for the unpolarized electroproduction cross
section is, after dividing out the flux of virtual photons,
\begin{equation}
d\sigma/d\Omega_K = \sigma_u + \epsilon \sigma_{TT} \cos(2\phi) +
\sqrt{2\epsilon(\epsilon+1)} \sigma_{LT} \cos(\phi),
\end{equation}
where $\sigma_U = \sigma_T + \epsilon \sigma_L$ is the 
unseparated sum of the transverse and longitudinal cross sections,
$\epsilon$ is the photon polarization parameter determined by the
electron scattering kinematics, and $\phi$ is the azimuthal angle
between the electron scattering and the hadronic reaction planes.  In
CLAS it is possible to detect the scattered electrons together with
the produced kaons over a broad range of kinematics to separate out
the $\phi$ angle dependence of the reaction, and thus determine
$\sigma_{U}$, $\sigma_{LT}$, and $\sigma_{TT}$.  A sample
result~\cite{feuerbach} is shown in Fig.~\ref{fig:c}, where three
components of the cross section are plotted as a function of $W$ for
both hyperons.  The data and model calculations have been averaged
over $0.5 < Q^2 <0.9$ GeV$^2$, and also averaged over the `backward'
kaon center-of-mass angles of $-0.67 < \cos(\theta_K) < -0.33$.

A major reason the cross section components for $\Lambda$ and
$\Sigma^0$ are different in Fig.~\ref{fig:c} is due to the isospin
selectivity, {\it i.e.} the $K^+\Lambda$ final state couples only to
$N^*$'s, while $K^+\Sigma^0$'s couple to both $N^*$'s and
$\Delta$'s. One sees that $\sigma_U$ in the $\Lambda$ cross section
has two peaks, as was seen in the photoproduction data, thus
confirming the strong contribution of a resonance-like structure near
1.9 GeV as seen in the photoproductin data.  In the $\Sigma^0$ case
there is only a broad peak without clear structure.  In the
interference cross sections the most prominent feature at these
kinematics is a broad negative dip in $\sigma_{TT}$ for the $\Sigma^0$
case.  One notes immediately that none of the models (the same models
cited previously) are able to consistently describe these results.
However, there are several well-established $\Delta^*$ resonances near
1.9 GeV which can couple to $K^+\Sigma^0$.  This means that more work
needs to be done to model resonance content and the $Q^2$ dependence
of the reaction mechanisms.

\begin{figure}[ht]
\resizebox{0.35\textwidth}{!}{\includegraphics{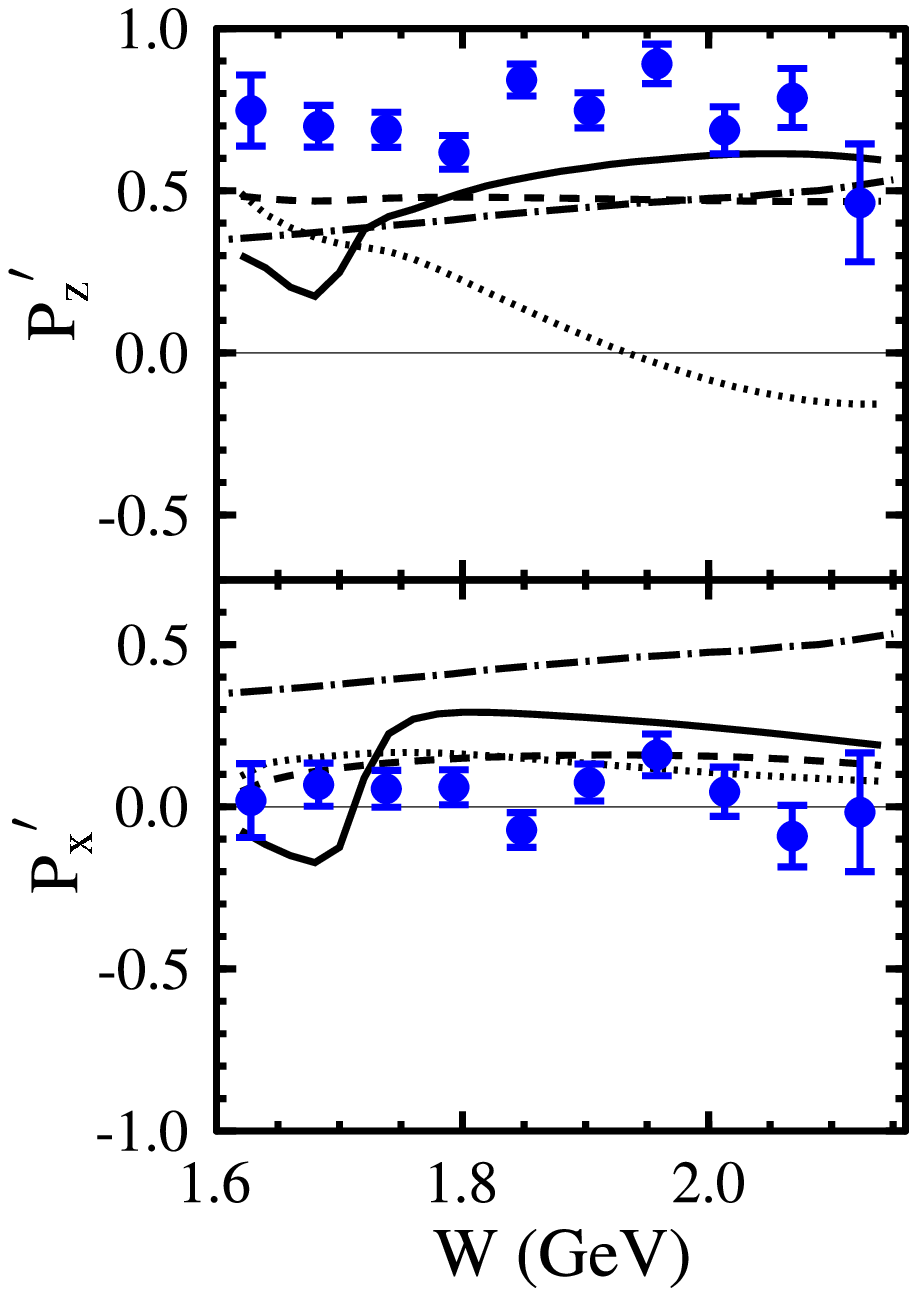}}   
\caption{Polarization transfer to the recoiling hyperon in the
reaction $\vec{e} + p \rightarrow e' + K^+ + \vec{\Lambda}$.  Top
panel shows the large $\Lambda$ polarization along the photon
direction, bottom panel shows the vanishing polarization transverse to
this direction. From Carman {\it et al.}~\cite{carman}.
\label{fig:d}}
\end{figure}

The circular polarization of virtual photons can be transferred to the
produced $\Lambda$'s in electroproduction, and this process has been
studied at CLAS~\cite{carman} using a $\sim70\%$ polarized electron
beam.  The ``beam-recoil'' double polarization thus measured is again
amenable to comparison with a variety of reaction models.  The two
components of $\Lambda$ polarization in the $K\Lambda$ plane can be
determined by measuring the beam-helicity-dependent yield asymmetry of
protons from the decay of the $\Lambda$'s.  The main result is that
the polarization of the $\Lambda$'s is maximal in the direction of the
virtual photon, and minimal transverse to this direction.
Fig.~\ref{fig:d} shows the data.  This phenomenon can be qualitatively
understood~\cite{carman} at the quark level in terms of the $s\bar{s}$
quark pair being formed predominantly in the anti-aligned 
state, with the produced $s$-quark carrying the $\Lambda$ spin,
polarized along the photon helicity direction.  It may also be a
manifestation of the VDM concept that the photon is first converted
into an aligned virtual $s\bar{s}$ pair in the form of a virtual
$\phi$ meson followed by quark rearrangement.

\begin{figure}[ht]
\begin{tabular}{cc}
\resizebox{0.35\textwidth}{!}{\includegraphics{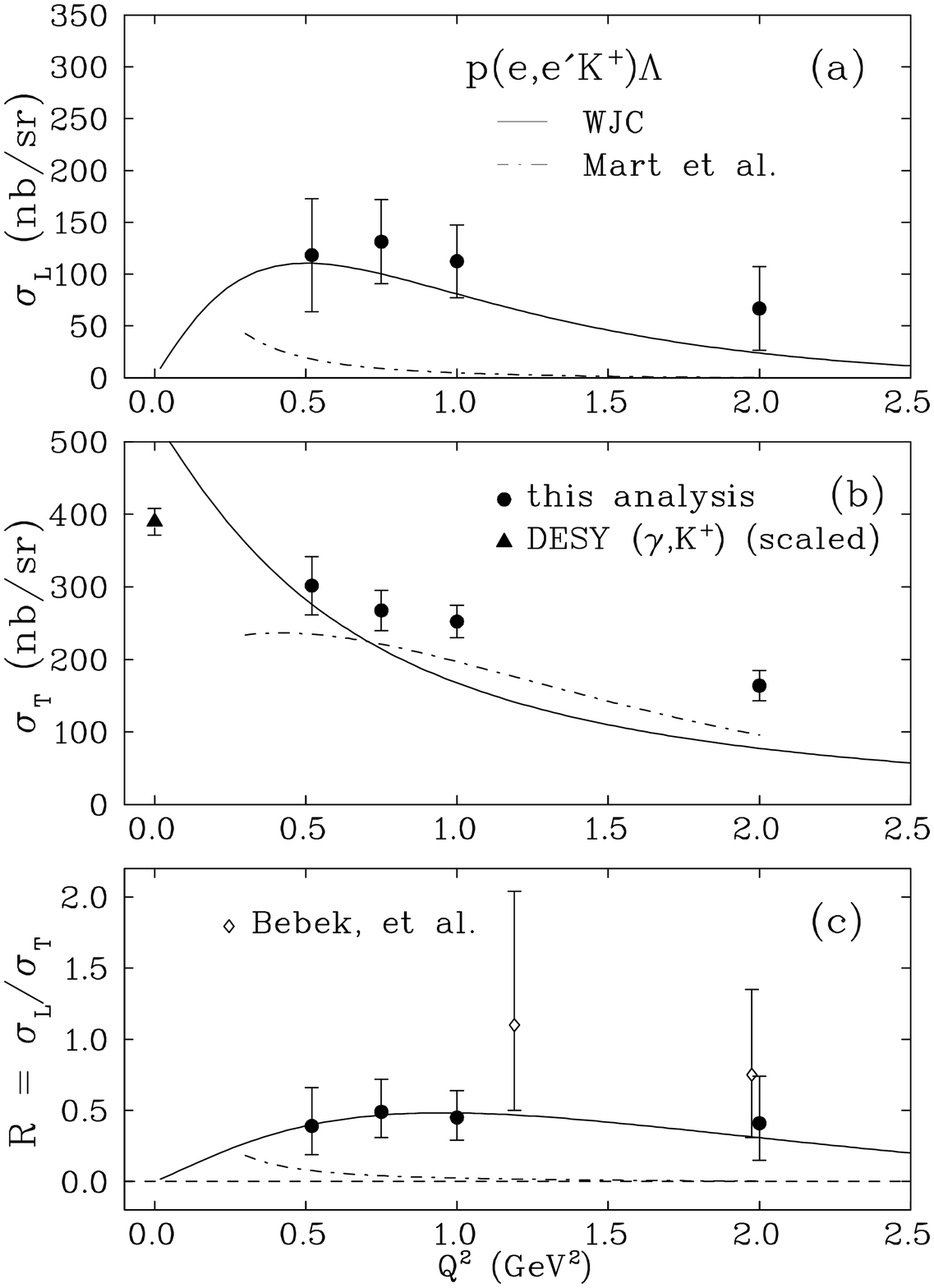}} &   
\resizebox{0.35\textwidth}{!}{\includegraphics{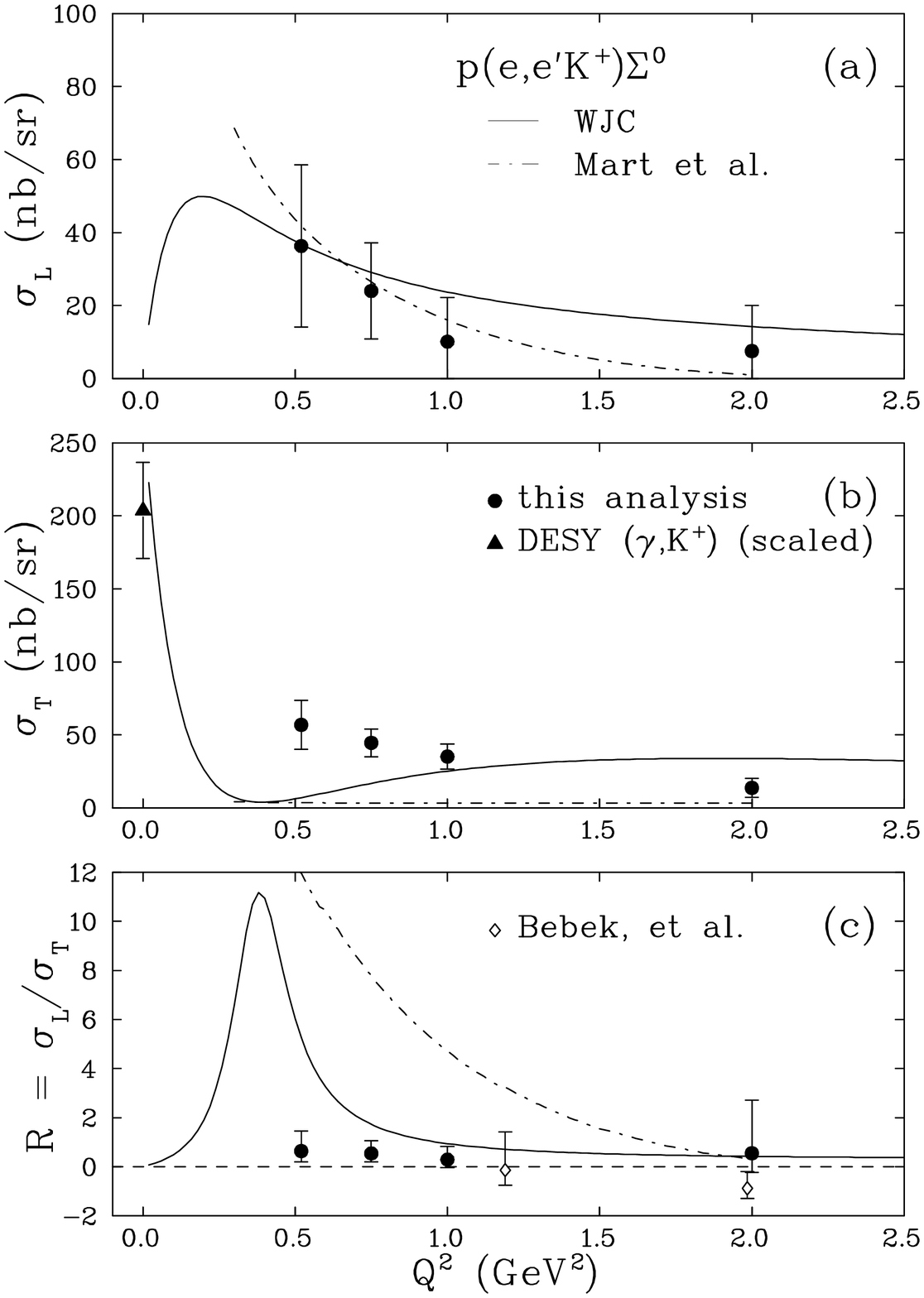}}   
\end{tabular}
\caption{
Results from Mohring {\it et al.}~\cite{mohring} for Rosenbluth separation of 
$\sigma_L$ (a), $\sigma_T$ (b), and their ratio (c).  The data are
compared to calculations of Refs.~\cite{wjc} and ~\cite{mart}.
Results for the $\Lambda$ are at left and $\Sigma^0$ at right.
\label{fig:e}}
\end{figure}

For the reactions $p(e,e'K^+)Y$ it has also been possible to make a
separation of the pieces of $\sigma_U$ into transverse and
longitudinal cross section using the traditional Rosenbluth method.
This has been done in Hall C at Jefferson Lab using the ``Short Orbit
Spectrometer'' to detect kaons in coincidence with the ``High Momentum
Spectrometer'' to detect scattered electrons~\cite{mohring}.  The
acceptance of these magnetic spectrometers led to a very narrow set of
kinematics at which the separation was made: specifically, the kaon
was along the direction of the virtual photon, {\it i.e.}
$\theta_{\gamma K} = 0^\circ$, and for $W=1.84$~GeV.  Fig.~\ref{fig:e}
shows the resulting cross section components and their ratio, for
$\Lambda$'s and $\Sigma^0$'s, as a function of $Q^2$.  These are the
most precise separations ever measured for these hyperons.  They show
that the ratio $\sigma_L/\sigma_T$ is about one half at these
kinematics up to $Q^2 = 2 $~GeV$^2$.

The calculations shown for comparison exhibit, at best, only fair
agreement with the data.  The work of Williams, Ji, and Cotanch
~\cite{wjc} is perhaps the better of the two, at least for the ratio
of $\sigma_L/\sigma_T$ for the $\Lambda$.  The authors of the
paper~\cite{mohring} surmise that the baryon resonance content of the
models, as well as the modeling of the hadronic form factors must be
improved in order to successfully reproduce these results.  This is
the same conclusion that can be reached from examination of the CLAS
data shown above.  Generally, all of these new high-precision data for
photo- and electro-production of strangeness on the proton call out
for a renewed effort at understanding baryon and meson exchange
structure of these reactions.

It would be very interesting to measure the charge form factor of the
$K^+$ meson in order to compare it with the form factor of the
$\pi^+$.  The same Hall C data set used in the analysis of the Mohring
{\it et al.}  result discussed above has been used by the E93-018
Collaboration~\cite{baker} to extrapolate the $t$-dependence of
$\sigma_L$ to the kaon pole (the Chew-Low method).  A single
unpublished form factor value $Q^2=1.0$ GeV$^2$ has been
obtained. Further analysis of data for values at $Q^2 = 0.5$ and $2.0$
GeV$^2$, taken in Halls C and A, is in progress in order to determine
the trend of this form factor.

The decay of the electroproduced excited hyperon $\Lambda(1520)
\rightarrow K^-p$ was studied at CLAS~\cite{barrow}. The angular
distribution in the $t$-channel helicity frame gives information about
whether the $t$-channel exchange particle had spin $J=0$ (such as the
$K^-$) or had $J$ greater than 0 (such as the $K^{-*}$). This is
interesting because in photoproduction, at $Q^2 = 0$, there is a clear
preference for $\Lambda(1520)$ formation via $K^{0*}$ exchange.  At
$Q^2 > 0.9$ GeV$^2$, the CLAS experiment showed that $J=0$ exchange
provides about $60\%$ of the formation channel.  These results are
striking, but have not yet lead to renewed theoretical activity on
this reaction mechanism.

\section{Production from Nuclei}
The Jefferson Lab data on hyperon electroproduction on the deuteron
and heavier nuclei have recently started to appear in archival
journals.  Other results have been shown at conferences only.  Below
we give an overview of the most interesting new results.  They come
from Hall C work, but there is also program getting underway in Hall
A.

\subsection{The $YN$ Interaction}\label{subsec:yn}
Quasi-free kaon production on the deuteron, $d(e,e'K^+)YN$, has been
measured by the E91-016 Collaboration in Hall C~\cite{reinhold}.  The
significance of this work is two-fold: it gives unique access to one
of the elementary production reactions, $n(e,e'K^+)\Sigma^-$, and
secondly, on one side of the quasi-free peak there is a kinematic
regime where the hyperon and the nucleon have nearly the same
momentum, and therefore interact via the low energy $YN$ interaction as
a final state interaction (FSI).

\begin{figure}[ht]
\resizebox{0.50\textwidth}{!}{\includegraphics{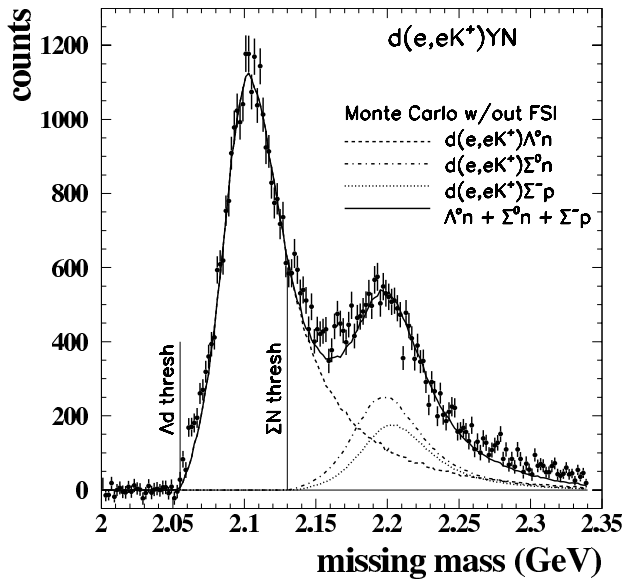}}   
\caption{
Reconstructed missing mass from $d(e,e'K^+)$ from Ref~\cite{reinhold}.
The curves are Monte Carlo simulations which do not have $YN$ final
state interactions turned on. FSI effects are seen as excess counts
near the $\Lambda d$ threshold.
\label{fig:f}}
\end{figure}

Figure~\ref{fig:f} shows the reaction at $Q^2 = 0.38$ GeV$^2$ and a
beam energy of 3.2 GeV.  The curves are Monte Carlo calculations which
do not include FSI.  The contribution from the neutron to $\Sigma^-$
production is substantial.  On the left-hand edge one sees the deficit
in the prediction of the quasi-free Monte Carlo model which must be
filled in by FSI.  Several $YN$ models have been used to compute the
final state interaction in the simulations (not shown here), each with
a characteristic S-wave scattering length and effective range; some
sensitivity has been shown, favoring the ``Verma'' and ``J\"{u}lich
A'' potentials and not favoring ``J\"{u}lich B''.

\subsection{Hypernuclear Electroproduction}\label{subsec:hyper}
A remarkable achievement of the Jefferson Lab program has been the
observation of electromagnetically produced hypernuclear states  in
light nuclei. Experiments using two different approaches have each
succeeded in finding evidence for such states.  

\begin{figure}[ht]
\resizebox{0.40\textwidth}{!}{\includegraphics{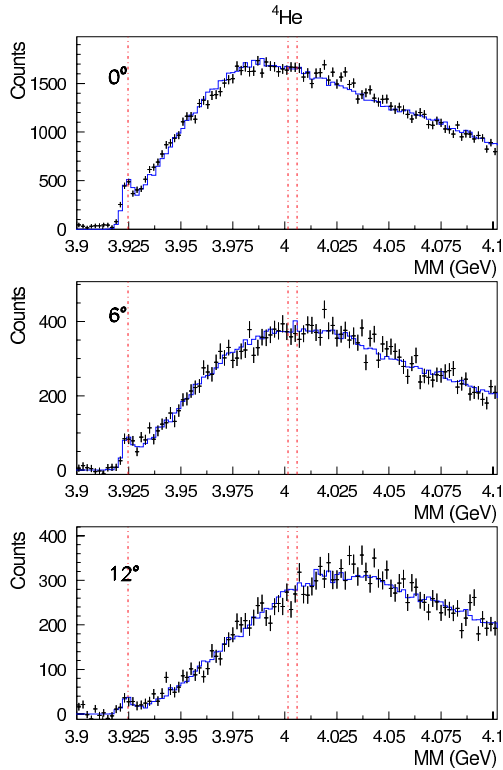}}   
\caption{
Missing mass distributions for $^4He(e,e'K^+)$ showing quasi-free Lambda
production and the formation of the bound state $_\Lambda^4H$, from 
Dohrmann {\it et al.}~\cite{dohrmann}. 
\label{fig:g}}
\end{figure}

Using the standard Hall C spectrometers, the E91-016
Collaboration~\cite{dohrmann} measured the reactions $A(e,e'K^+)YX$
for $^1H, ^2H, ^3He, ^4He, C,$ and $Al$ targets, at $Q^2=0.35$ GeV$^2$
and for the lab angle between the virtual photon and the kaon at $0^o,
6^o,$ and $12^o$.  The most compelling result was for the $^4He$
target, shown in Fig.~\ref{fig:g} which showed clear bound state peaks
at the end of the quasi-free production spectrum for all three
scattering angles.  These peaks correspond to the formation of the
well-known hypernucleus $_\Lambda^4H$.  The cross section was
estimated to be about 20 nb/sr.  The data in the figure shows no
evidence of $\Sigma$ hypernuclear production (dashed lines high in
quasi-free spectrum).

The limitation of the experimental technique used to produce this
result is that the missing-mass resolution is insufficient to extend
these measurements to heavier hypernuclei where many excited states
are present.  Another method which achieves resolution of under 1 MeV
would be required to make electromagnetic hypernuclear production
truly interesting as a research tool for hypernuclear spectroscopy.

\begin{figure}[ht]
\resizebox{0.50\textwidth}{!}{\includegraphics{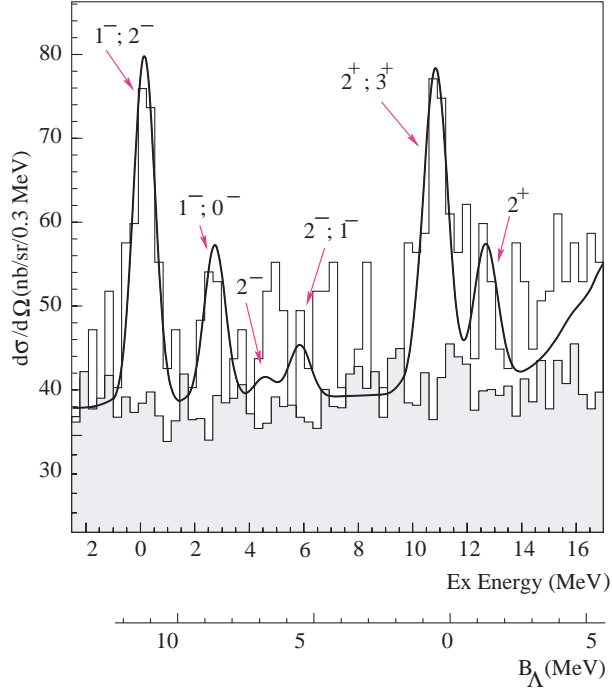}}   
\caption{
Missing mass distributions for $^{12}C(e,e'K^+)^{12}_\Lambda B$ showing peaks
corresponding to the ground state ($\Lambda$ in the S shell), and P-shell
excited state ($\Lambda$ in the P shell).  The curve shows predictions
for some core-excited states in which the host $^{11}B$ nucleus is not in its
ground state. From Ref~\cite{miyoshi}. 
\label{fig:h}}
\end{figure}

This other method was pioneered by the E89-009
collaboration~\cite{miyoshi}.  They used the Hall C SOS spectrometer to
detect kaons, but used a dedicated dipole ``Enge splitpole''
spectrometer for the electrons.  In addition, a beam-splitter magnet
was used to allow the electron angle to be pushed all the way to zero
degrees, thus decreasing $Q^2$ and maximizing the rate of good events.
Also, great care was taken to minimize straggling effects in
materials.  The result was obtained~\cite{miyoshi} for the reaction
$^{12}C(e,e'K^+)^{12}_\Lambda B$, seen in Fig.~\ref{fig:h}, which
showed peaks corresponding to bound $\Lambda$'s in the S and in the
P shells.

The experiment achieved a missing mass resolution of 0.9 MeV {\small FWHM},
which can be compared with about 1.5 MeV as the best value achieved
using the $(\pi^+,K^+)$ reaction for hypernuclear spectroscopy.  The
experiment was limited by statistics and resolution, so that little
additional quantitative information has emerged.  The experiment
demonstrated that spectroscopic resolution is achievable, but to turn
the method into a research tool a better instrument is needed.  A new
kaon spectrometer has been built~\cite{fujii} that should achieve 0.4
MeV resolution and permit count rates of over several 100 per day for
the $^{12}_\Lambda B$ ground state.  It will be installed at Jefferson Lab
in mid-2004.

In Hall A at Jefferson Lab the E94-107 Collaboration~\cite{markowitz}
is also planning to start hypernuclear spectroscopy in late 2003.
They will use the existing HRS spectrometers, augmented by a
septum-magnet arrangement which will allow the electron spectrometer
to reach $6^o$ in the lab and by a RICH detector for more effective
$K/\pi$ separation.

\section{The $\Theta^+$ state}
In the naive quark model, baryons come only in colorless 3-quark
flavor combinations. However, no fundamental rule forbids combinations
of more quarks as long as they are also colorless.  Hence, for decades
there have been searches for other multi-quark structures that would,
if they exist, be important ingredients of hadronic physics.  No such
unusual states had been convincingly identified however, so the Fall
2002 announcement by the LEPS group at SPring-8~\cite{nakano} in Japan
of a narrow pentaquark state near 1540 MeV that decays into $K^+n$
caused renewed excitement.  This state is manifestly ``exotic'' since
the $K^+$ contains an anti-$s$ quark, which cannot be an ingredient in
an ordinary baryon.  A large number of theoretical papers have
appeared in recent months, as well as at least 4 experimental
confirmations~\cite{nakano,itep,stepanyan,barth}, showing that this
topic has become very hot.  Spring-8's LEPS group was inspired by the
theoretical prediction of Diakonov, Petrov, and
Polyakov~\cite{diakonov} who predicted a narrow $uudd\bar{s}$
pentaquark state within the framework of a chiral soliton model, in
which it appears as one member of an anti-decuplet of 5-quark states.
The observation of a state with some of the predicted features is not
yet final proof, of course, that this model is the optimal explanation.

\begin{figure}[ht]
\resizebox{0.8\textwidth}{!}{\includegraphics{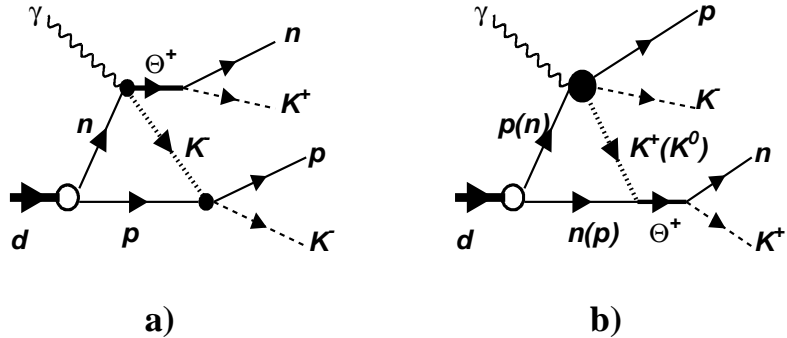}}   
\vspace{-1.0in}
\caption{
Diagrams leading to the production and decay of a $\Theta^+$ from 
exclusive reactions on the deuteron.
\label{fig:i}}
\end{figure}

The CLAS Collaboration at Jefferson Lab has recently
reported~\cite{stepanyan} results from the exclusive reaction $\gamma
d \rightarrow K^+ K^- p n$ in which the neutron is reconstructed via
the missing mass off the three detected charged particles.  Two of
many possible diagrams through which the $\Theta^+$ might be formed are
shown in Fig.~\ref{fig:i}.  The ``two-step'' nature of the production
mechanism has the experimental benefit that the $K^-$'s that are
typically very forward peaked rescatter to larger lab angles, and
therefore are more inside the CLAS acceptance.

\begin{figure}[ht]
\resizebox{0.50\textwidth}{!}{\includegraphics{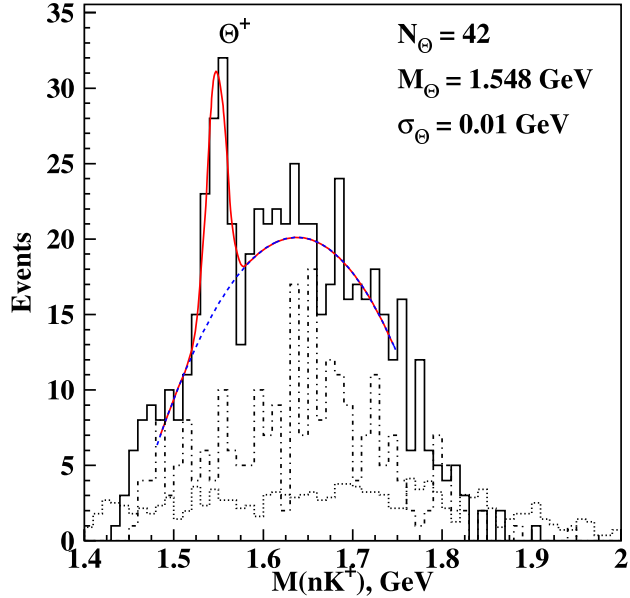}}   
\caption{
Invariant mass of the $K^+n$ system, which has strangeness $S = +1$,
showing a sharp peak at the mass of 1.548 (GeV/c)$^2$.  The fit (solid
line) to the peak on top of the smooth background (dashed line) gives
a statistical significance of about 5 $\sigma$.  The dashed histogram
shows events removed by the hyperon cut, and the dotted histogram
shows the scaled background from non-pKK events.
\label{fig:j}}
\end{figure}

The $\Theta^+$ signal was sought in the invariant mass of the $K^+n$
decay channel, and the final result is shown in Fig.~\ref{fig:j}.  The
signal is deduced to have about ``5 sigma'' significance, using a
smooth (polynomial) background.  Various background shapes have been
tested, with similar significances.  Production of excited hyperons
is present in the same final state, and forms background under the
peak in the final spectrum; the $\Lambda(1520)$ events were removed
explicitly, however, with a cut on the $K^-p$ mass distribution.  Full
Monte Carlo modeling of the background has not been completed,
although preliminary tests have suggested that the background has a
large contribution from non-resonant $KK$ production off the bound
nucleon.

Immediately one wonders whether the $\Theta^+$ state could also be
reached in photoproduction off the proton directly.  At CLAS there is
a search at this time for the reaction $\gamma + p \rightarrow
\bar{K^0} + \Theta^+ \rightarrow \pi^+ \pi^- K^+ n$.  A positive
result has been reported recently from SAPHIR~\cite{barth}, in the
same channel.  Interestingly, CLAS does have evidence for the
$\Theta^+$ in the reaction $\gamma + p \rightarrow K^- + \pi^+ +
\Theta^+ \rightarrow K^- + \pi^+ + K^+ + n$.  (The result was shown in
the oral presentation but is not included here since the data are
still in flux.)  In conclusion, the study of this new exotic baryonic
state is very intense at this time, with positive sighting of a narrow
$\Theta^+$ by at least 4 experimental groups, and with a plethora of
theoretical interpretation underway.

\section{Other Works in Progress}

Additional work in progress on strangeness production at Jefferson Lab
has not been mentioned so far due to time and space constraints.  In
the future we hope to present results on the following topics: (i) the
radiative decay of the $\Sigma$(1385), which tests quark-model
wave-functions~\cite{mutchler}, (ii) cross sections~\cite{carnahan}
for $\gamma p \rightarrow \bar{K^0}\Sigma^+$ (iii) the beam-recoil
double polarization observables $C_x^\prime$ and $C_z^\prime$ in
$p(\gamma,K^+)\Lambda$~\cite{bradford}, (iv) cross sections for $K^+$
photoproduction off the deuteron~\cite{ioana}, (v) results for the
``fifth'' structure function in electroproduction~\cite{raue}, (vi)
photoproduction cross sections for the excited
hyperons~\cite{juengst}, (vii) line-shape analysis for photoproduction
of the $\Lambda(1405)$ decaying to different $\Sigma\pi$ charge
states, as a test of its structure~\cite{ras}; (viii) photoproduction
of $S=-2$ Cascade ($\Xi^-$) states~\cite{price}.

\section{Summary}
The disparate experimental results presented here make it difficult
to summarize this report in a few words or a single central new
concept.  The main observations are that (a) interesting new baryon
resonance structure has been seen in both photo- and electro-production
of strangeness on the nucleon; (b) electroproduction off the deuteron
has shown that it is possible to test models of the $YN$ final state
interaction; (c) hypernuclear production from $^3He$ and $^4He$ and
from $^{12}C$ shows the feasibility of using electromagnetic probes to study 
production and structure of hypernuclei; and (d) an exotic new
baryon, the $\Theta^+$, has been confirmed to exist at Jefferson Lab,
a discovery which is engendering a great deal of new theoretical
and experimental work.

\section*{Acknowledgments}
The author wishes to thank numerous colleagues who made their results
available for this talk and paper.  They include K. Baker, S. Barrow,
E. Beise, D. Carman, F. Dohrmann, K. Hicks, J. Melone, R. Feuerbach,
V. Koubarovski, K. Livingston, P. Markowitz, J. McNabb, M. Mestayer,
G. Niculescu, J. Price, J. Reinhold, S. Stepanyan, and L. Tang.  They
and other colleagues offered advice and suggestions which are
gratefully acknowledged. This work is supported by DOE contract
DE-FG02-87ER40315.


\begin{thebibliography}{0}

\bibitem{clas0} B. Mecking {\it et al.}, Nucl. Instrum. and Methods 
                {\bf A503}, 513 (2003), and references therein. 

\bibitem{mcnabb} J.W.C. NcNabb  \emph{et al} 
{\it nucl-ex/0305028}; submitted to Phys. Rev. Lett.; 
W. J. C. McNabb Ph.D. Thesis, Carnegie Mellon 
University (2002) (unpublished). 

\bibitem{maid} T. Mart, C. Bennhold, H. Haberzettl,
and L. Tiator, ``KaonMAID 2000'' at 
www.kph.uni-mainz.de/MAID/kaon/kaonmaid.html.

\bibitem{mart} T. Mart and C. Bennhold, Phys. Rev. {\bf C61}, 012201 (2000);
C. Bennhold, H. Haberzettl, and T. Mart, 
{\it nucl-th/9909022} and 
Proceedings of the 2nd Int'l Conf. on Perspectives in Hadronic
Physics, Trieste, S. Boffi, ed.,  World Scientific, (1999).

\bibitem{jan} S. Janssen, J. Ryckebusch, D. Debruyne, and T. Van
  Cauteren, Phys. Rev {\bf C65}, 015201 (2001); 
  S. Janssen {\it et al.}, 
  Eur. Phys. J. {\bf A 11}, 105 (2001); curves via private communication. 

\bibitem{laget} M. Guidal, J.-M. Laget, and M. Vanderhaeghen,
 Phys Rev. {\bf C61}, 025204 (2000);  M. Vanderhaeghen, M. Guidal, 
and J.-M. Laget, Phys. Rev. {\bf C57}, 1454 (1998).

\bibitem{bonn} M. Q. Tran {\it et al.}, Phys. Lett. {\bf B445}, 20 (1998);
M. Bockhorst {\it et al.}, Z. Phys. {\bf C63}, 37 (1994).

\bibitem{klein} CLAS Experiment 98-109, P. Cole, spokesperson, and
K. Livingston and J. Melone, (private communication). 

\bibitem{bradford} CLAS Experiment 89-004, R. Bradford, PhD thesis.

\bibitem{feuerbach} R. Feuerbach \emph{et al.}, 
to be published. Ph.D. Thesis, Carnegie Mellon 
University (2002) (unpublished).  

\bibitem{carman} D. S. Carman \emph{et al.}, {\it Phys. Rev. Lett.}
{\bf 90}, 131804 (2003); M. Mestayer, these proceedings.

\bibitem{mohring} R. M. Mohring \emph{et al.}, {\it Phys. Rev.}
{\bf C67}, 055205 (2003).  See also the earlier paper for the same
experiment: G. Niculescu \emph{et al.}, {\it Phys. Rev. Lett.}
{\bf 81}, 1805 (1998).

\bibitem{wjc} R.A.Williams, C.-R. Ji, and S. R. Cotanch {\it Phys. Rev.}
{\bf C46}, 1617 (1992).

\bibitem{baker} O. K. Baker for the E93-018 collaboration (private communication).

\bibitem{barrow} S. P. Barrow \emph{et al.}, {\it Phys. Rev.}
{\bf C64}, 044601 (2001).

\bibitem{reinhold} J. Reinhold for the E91-016 collaboration, 
  Proceedings of Baryons 2002, C. Carlson and B. Mecking, Eds., World
  Scientific, 589, (2002),  and private communication.

\bibitem{dohrmann} F. Dohrmann for the E91-016 collaboration,
  Proceedings of Baryons 2002, C. Carlson and B. Mecking, Eds., World
  Scientific, 585, (2002),  and private communication.

\bibitem{miyoshi} T. Miyoshi \emph{et al.}, {\it Phys. Rev. Lett.}
{\bf 90}, 232502 (2003).

\bibitem{fujii} Y. Fujii  \emph{et al.}, {\it Nucl. Phys.}
{\bf A721}, 1079c (2003).

\bibitem{markowitz} P. Markowitz for the E94-107 collaboration (private communication).

\bibitem{nakano} T. Nakano \emph{et al.}, {\it Phys. Rev. Lett.}
{\bf 91}, 092001 (2003).

\bibitem{itep} V. V. Barmin \emph{et al.}, hep-ex/0304040, (2003);
  accepted by Yad. Phys.

\bibitem{stepanyan} S. Stepanyan \emph{et al.}, hep-ex/0307018, (2003);
 submitted to {\it Phys. Rev. Lett.}

\bibitem{barth} J. Barth \emph{et al.}, hep-ex/0307083, (2003).

\bibitem{diakonov} D. Diakonov, V. Petrov, and M. Polyakov, {\it Z. Phys.}
{\bf A359}, 305 (1997).

\bibitem{mutchler} CLAS Experiment 89-024, G. Mutchler, spokesman. 
\bibitem{carnahan} CLAS Experiment 89-004, B. Carnahan, Ph.D. Thesis,
Catholic Univ. of America. 
\bibitem{ioana} CLAS Experiment 89-045, B. Mecking, spokesman, I. Niculescu, contactperson. 
\bibitem{raue} CLAS Experiment 99-006, D. Carman and B. Raue, spokesmen. 
\bibitem{juengst} CLAS Experiment 89-004, H. Juengst, contactperson. 
\bibitem{ras} CLAS Experiment 89-004, R. Schumacher, spokesman. 
\bibitem{price} CLAS Approved Analysis, J. Price, spokesman. 

\end{thebibliography}
\end{document}